\documentclass[twocolumn]{aastex631}
\usepackage{amsmath,amssymb}
\usepackage{graphicx}
\graphicspath{{figures/}}
\usepackage{booktabs}
\usepackage{xcolor}
\usepackage{url}

\shorttitle{A Source--Gate--Transfer Test for PDS}
\shortauthors{Podladchikova}
\newcommand{\rsun}{R_{\odot}}
\newcommand{\kms}{\mathrm{km\ s^{-1}}}
\newcommand{\tb}{tB}
\newcommand{\pb}{pB}
\newcommand{\praw}{p_{\mathrm{raw}}}
\newcommand{\aeff}{A_{\mathrm{eff}}}

\begin{document}
\title{Where Do Periodic Density Structures Acquire Coherence?\\
A Low-Coronal Resonator Candidate and a 12-Event STEREO/SECCHI Transfer Test}
\author{Olena Podladchikova}
\affiliation{National Technical University of Ukraine, Igor Sikorsky Kyiv Polytechnic Institute,
37 Beresteiskyi Prospect, Kyiv 03056, Ukraine}
\affiliation{Leibniz-Institut f\"ur Astrophysik Potsdam (AIP),
An der Sternwarte 16, 14482 Potsdam, Germany}
\correspondingauthor{Olena Podladchikova}
\email{epodlad@gmail.com}

\begin{abstract}

Periodic density structures (PDS) are trains of plasma enhancements carried into the heliosphere, but their repeated timing does not by itself identify where the cadence is set. We test whether PDS selected independently in the outer corona can be traced back to a low-coronal clock, and whether that organization survives through the EUVI--COR1--COR2 observing chain.

We analyze 12 STEREO-A/SECCHI events on 2008 January 11--14 along a nonradial path. EUVI 171~\AA\ at $1.10$--$1.20\,\rsun$ shows an unusual concentration of power in the predeclared 80--130 minute band relative to filter-matched red-noise controls, while the ensemble is most strongly organized in upper COR1 at $2.5$--$3.0\,\rsun$. These signatures identify a low-coronal modulation candidate and an intermediate-height organization domain, but they do not form a unique phase-preserving clock extending into COR2.

Instead, the more persistent observable is spatial order. Events 9 and 12 retain expansion-stable outward ordering and contain four and three radial nodes, respectively. In polar $r$--$\theta$ maps, intersections of oppositely inclined, expansion-aware matched-filter ridge supports form repeated X-/diamond-like patterns. Their outer nodes enter the upper-COR1 organization domain. The morphology is compatible with stationary or quasi-stationary shock-cell processing, although the brightness diagnostics are insufficient to establish a stationary MHD shock branch.

The observations therefore favor an intermittent source--gate--transfer picture: a low-coronal cadence may modulate plasma release, while event-dependent propagation progressively destroys exact phase coherence. The outward ordering of the density structures can survive, providing a more persistent signature of the source-to-wind transfer than phase locking itself.

\end{abstract}

\section{The Physical Question}\label{sec:intro}

Periodic density structures are routinely observed in the slow solar wind, but a repeated train does not uniquely identify its origin. The cadence may retain a low-coronal clock, arise intrinsically from reconnection or tearing, or be modified while the structures cross the young solar-wind acceleration region. STEREO/SECCHI observations already trace outward disturbances from EUVI through COR1 and COR2 \citep{Viall2010,Viall2015,Alzate2021,Alzate2023,Alzate2024}. Here we use that established nonradial path and event sequence to ask a different question: where does the observed organization become strongest, and how much of a possible source cadence survives into the outward density pattern?

This question connects two parts of solar-wind physics that are often studied
separately. The photosphere and convection zone continually restructure the
magnetic boundary conditions and supply stresses, waves, and flux to the
corona, but the escaping density pattern need not preserve that forcing
one-to-one. Between the lower corona and COR2, plasma crosses the streamer
cusp/current-sheet environment and the principal acceleration region of the
young solar wind. Release, reconnection, expansion, and MHD processing can each
alter the visibility, separation, and phase of successive density structures.
The observational problem is therefore not simply to find a period, but to
determine which property of a PDS train---cadence, phase, order, width, or
contrast---survives each stage.

Open--closed magnetic boundaries, the S-Web, streamer cusps, and the heliospheric current sheet provide plausible release environments \citep{Antiochos2011,Reville2020}; streamers can support MHD responses, while tearing and plasmoid formation can make reconnection intrinsically repetitive. The experiment below tests which, if any, of these physical elements is visible in this event sequence.

Coronal streamers are not passive brightness backgrounds. For example,
\citet{Feng2013} reported a 70--80 minute kink-like streamer oscillation whose
wavelength evolved between approximately $2$ and $3\,\rsun$, interpreting it
as a probable streaming kink-mode Kelvin--Helmholtz instability. That event is
not evidence for the present 80--130 minute candidate, but it demonstrates that
streamer/current-sheet systems can support observable MHD organization on
comparable macroscopic timescales. The relevant question here is whether such
dynamics merely modulate release, or whether their phase remains identifiable
in the later advected density train.

Two earlier arXiv preprints motivate the present experiment \citep{Podladchikova2025a,Podladchikova2025b}. The first separates the plasma source from a possible MHD modulator: a compact slow response, a surface or current-sheet disturbance, or a transient kink response can modulate reconnection without transporting a density packet through closed field. The second explored a broader resonator--Laval-nozzle interpretation. The analytical development presented there shows that a stationary advective flow can strongly accelerate and spatially stretch a train while preserving its temporal cadence. The present paper does not treat those hypotheses as observationally established; it converts them into separable tests. Together they suggest a four-part system:
\begin{equation*}
\begin{aligned}
\boxed{\text{driver}}&\longrightarrow\boxed{\text{MHD response}}\\
&\longrightarrow\boxed{\text{reconnection gate}}
\longrightarrow\boxed{\text{wind transfer}} .
\end{aligned}
\end{equation*}
Here the \emph{clock} is a recurring process that sets or modulates release cadence; it is not the PDS train itself. A low-coronal MHD response is one candidate, while intrinsic reconnection can provide another.

The central question is:
\begin{quote}
\textit{Do PDS reveal a universal low-coronal clock, or do they acquire observable coherence only when an intermittent MHD gate releases and processes plasma into the solar wind?}
\end{quote}

The alternatives make different predictions. A source modulator should leave a phase-related signal below the cusp. A stationary transfer path should preserve relative timing even while acceleration changes spatial separation. A reforming gate can preserve outward order while adding event-dependent phase delay. Because the solar wind must cross a critical transition, we also test whether the transfer region is smooth or shows stationary shock-cell processing: the latter should add repeatable spatial nodes and, for a full MHD-shock identification, an independent plasma transition.

\begin{figure*}[!t]
\centering
\includegraphics[width=0.98\textwidth]{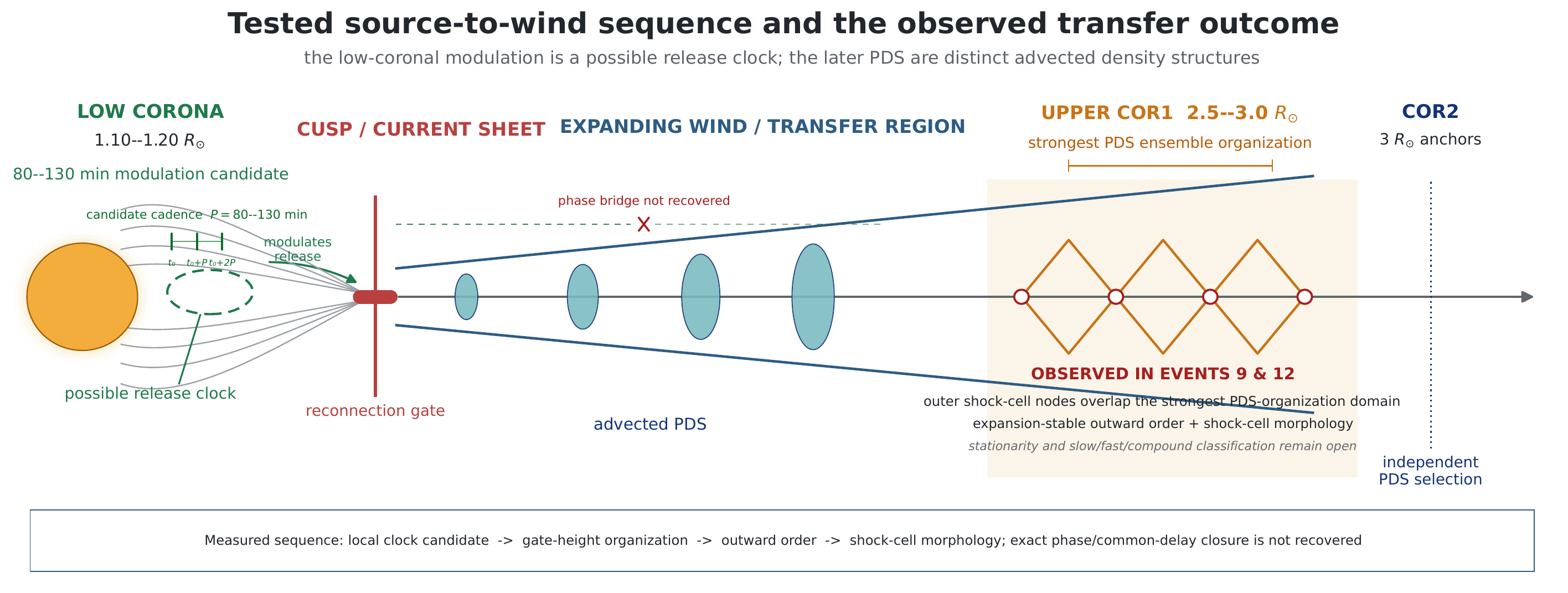}
\caption{Source--gate--transfer experiment and observed transfer outcome. The local EUVI 171~\AA\ 80--130 minute modulation candidate at $1.10$--$1.20\,\rsun$ is shown as a possible release clock; the cadence marks $t_0$, $t_0+P$, and $t_0+2P$ illustrate the tested periodicity rather than three identified releases. The broken phase connector indicates that a phase-preserving bridge from the low corona into COR1 is not recovered. Released density structures are advected through the expanding wind, while the strongest PDS ensemble organization appears in upper COR1 at $2.5$--$3.0\,\rsun$. Events 9 and 12 show expansion-stable outward order and shock-cell morphology in this domain, with their outer nodes overlapping the strongest organization region. The pattern is therefore drawn as possible processing of outward PDS in the transfer region, not as a second object launched after the PDS; node stationarity and slow/fast/compound MHD classification remain open.}
\label{fig:logic}
\end{figure*}

Figure~\ref{fig:logic} summarizes the tested sequence without identifying the modulation with the PDS themselves. A \emph{ridge} is an image-level brightness feature; it supports transport only when followed outward with compatible timing. Shock-cell morphology is a stronger geometric pattern, while a stationary MHD shock additionally requires upstream/downstream plasma-state closure.

The ordering in Figure~\ref{fig:logic} is spatial and diagnostic rather than a claim that a shock is created \emph{after} a PDS. The PDS are the advected density structures; the X-/cell-like pattern, where present, is a possible way those structures are processed while they cross the upper-coronal transfer region.

\section{What Each Physical Picture Predicts}\label{sec:physics}

\subsection{A clock is not the same object as a PDS}

PDS are primarily advected plasma structures, not standing waves. An MHD response can nevertheless modulate their release: a compressive slow response can perturb pressure and density, a surface/current-sheet response can change reconnection inflow, and a kink-like response can move the gate geometrically. The broad 80--130 minute interval is therefore tested as a physical band rather than as an exact line at 120 or 128.21 minutes. Compact slow-mode paths and flow-modified tearing can both produce comparable timescales \citep{Podladchikova2025a,Reville2020}; period agreement alone is not mode identification.

\subsection{Why a critical transition is unavoidable but a shock is not}

The solar wind must pass from a subcritical coronal state to an outflow that is supercritical with respect to the relevant characteristics. In a simple isothermal one-dimensional description,
\begin{equation}
\begin{aligned}
 \left(M^2-1\right)\frac{d\ln u}{ds}
      &= \frac{d\ln \aeff}{ds},\\
 \aeff(s)&=A(s)\exp[-\Phi_g(s)/c_s^2],
\end{aligned}
\label{eq:nozzle}
\end{equation}
where $s$ follows the flow, $u$ is speed, and $M=u/c_s$. A smooth sonic transition is possible at an extremum of $\aeff$; super-radial expansion can create multiple mathematical critical points \citep{KoppHolzer1976,HabbalTsinganos1983}. A discontinuous shock is required only if the boundary conditions cannot be connected by an admissible smooth branch, and multiple-critical-point models can admit standing inner-wind shocks \citep{Habbal1985,Li2011}.

\begin{figure}[!t]
\centering
\includegraphics[width=\columnwidth]{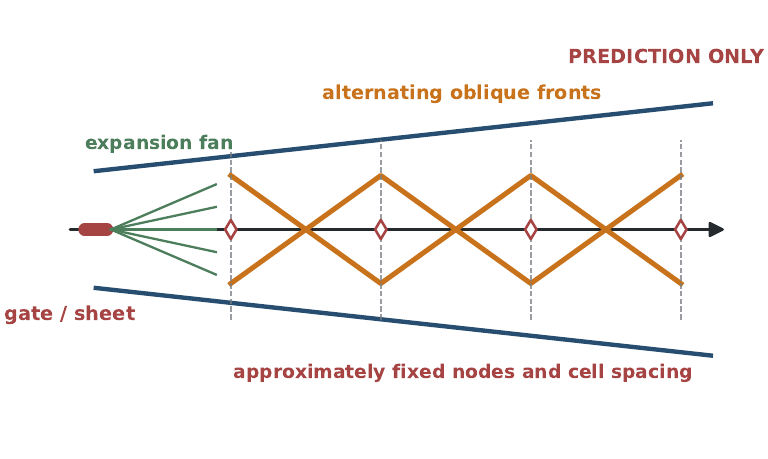}
\caption{Simulation-motivated stationary shock-cell signature. A finite-vertex expansion fan and oppositely inclined compression fronts can form diamond cells with approximately fixed nodes. The schematic shows the underexpanded ordering, in which an expansion fan appears first; an overexpanded jet instead begins with an inward-directed oblique shock. The reconnection-jet shock diamonds resolved by \citet{Zenitani2015} provide the MHD motivation for this test, and the schematic is conceptually redrawn from that mechanism. The morphology motivates the observational test; stationary MHD-shock identification additionally requires a plasma jump and characteristic-speed closure.}
\label{fig:shockprediction}
\end{figure}

In MHD, slow, Alfv\'en, and fast characteristics allow slow, fast, or compound transitions. Images alone cannot classify them: magnetic field, temperature, normal flow/shock speed, and a density or pressure jump are needed. Nevertheless, low-$\beta$ reconnection simulations produce repeated shock diamonds and expansion/compression cells in a Petschek outflow \citep{Zenitani2015}. Thus X-/diamond-like nodes and repeatable cell spacing are positive shock-cell morphology; a stationary MHD-shock claim remains the stronger plasma-transition test. Figure~\ref{fig:shockprediction} summarizes the simulation-motivated morphology used for this test.

\subsection{The timing theorem and its failure mode}

Let a source release structures at times $t_n^{(0)}$. If every structure follows the same path through a stationary flow, its arrival time at radius $r$ is
\begin{equation}
t_n(r)=t_n^{(0)}+\tau(r),
\qquad
\tau(r)=\int_{r_0}^{r}\frac{dr'}{u(r')}.
\label{eq:delay}
\end{equation}
The waiting time is therefore
\begin{equation}
P_n(r)=t_{n+1}(r)-t_n(r)=P_n^{(0)}.
\label{eq:cadence}
\end{equation}
Acceleration changes the spatial separation, $\lambda(r)=u(r)P_n$, but not the cadence. Even a fixed shock can preserve Equation~(\ref{eq:cadence}) if every pulse experiences the same shock state. Its signature must instead be an abrupt and repeatable change of density, width, temperature, or entropy.

If the gate or shock moves, each structure receives a different delay $\tau_n$:
\begin{equation}
P_n(r)=P_n^{(0)}+\tau_{n+1}(r)-\tau_n(r).
\label{eq:jitter}
\end{equation}
The added term is phase jitter. It can destroy exact speed closure while leaving the radial order recognizable. This is the signature that most directly separates a stationary nozzle from a reforming gate.

Table~\ref{tab:hypotheses} turns the physical alternatives into one ordered decision chain: first ask whether a source-side clock is present, then whether timing survives the gate and wind transfer, and only then whether the organized pattern reaches the stronger stationary-shock criteria.

\begin{table*}[!t]
\centering
\caption{Physical pictures, their decisive prediction, and what the present experiment actually returns. The rows follow the source-to-wind sequence used throughout the paper; passing one row does not imply the stronger rows below it.}
\label{tab:hypotheses}

{\fontsize{7.7}{8.7}\selectfont
\setlength{\tabcolsep}{3.2pt}
\renewcommand{\arraystretch}{1.12}
\begin{tabular}{lll}
\toprule
\parbox[t]{0.20\textwidth}{Physical picture} &
\parbox[t]{0.34\textwidth}{Decisive observational prediction} &
\parbox[t]{0.39\textwidth}{Present result} \\
\midrule
\parbox[t]{0.20\textwidth}{Low-coronal resonator / MHD modulator} &
\parbox[t]{0.34\textwidth}{A reproducible source-side signal whose phase is related to later releases.} &
\parbox[t]{0.39\textwidth}{Local EUVI 171~\AA\ 80--130 minute excess is present at $1.10$--$1.20\,\rsun$; the cross-height phase bridge is not recovered. \textbf{Candidate clock, not an identified mode or trigger.}} \\
\specialrule{0.18pt}{0.6pt}{0.6pt}
\parbox[t]{0.20\textwidth}{Intrinsic tearing / reconnection} &
\parbox[t]{0.34\textwidth}{Intermittent release can occur without one coherent upstream clock.} &
\parbox[t]{0.39\textwidth}{Not excluded. The missing universal phase relation leaves intrinsic reconnection as a viable source of event-to-event variability.} \\
\specialrule{0.18pt}{0.6pt}{0.6pt}
\parbox[t]{0.20\textwidth}{Smooth stationary transfer} &
\parbox[t]{0.34\textwidth}{Outward order and one common propagation delay should survive while the train stretches spatially.} &
\parbox[t]{0.39\textwidth}{Events 9 and 12 retain outward order, but no unique common-delay chain closes. \textbf{Order survives more robustly than phase.}} \\
\specialrule{0.18pt}{0.6pt}{0.6pt}
\parbox[t]{0.20\textwidth}{Reforming MHD gate / transfer region} &
\parbox[t]{0.34\textwidth}{Outward order may persist while delay, width, or vertex position changes between releases.} &
\parbox[t]{0.39\textwidth}{Mixed event classes plus loss of common-delay closure are consistent with intermittent or reforming processing near the cusp/current sheet.} \\
\specialrule{0.18pt}{0.6pt}{0.6pt}
\parbox[t]{0.20\textwidth}{Stationary shock-cell processing} &
\parbox[t]{0.34\textwidth}{Approximately fixed nodes/cell spacing plus an independent signed plasma transition and MHD closure.} &
\parbox[t]{0.39\textwidth}{Events 9 and 12 show shock-cell morphology and their outer nodes enter the upper-COR1 organization domain. \textbf{Stationarity and slow/fast/compound classification remain open.}} \\
\bottomrule
\end{tabular}
}
\end{table*}

\section{Observations and Measured Quantities}\label{sec:data}

\subsection{The 12-event release-clock and transfer test}

We use STEREO-A/SECCHI observations from 2008 January 11--14
\citep{Howard2008}. This is the path-14 interval in which
\citet{Alzate2024} reported clearly visible outward-propagating disturbances
across the combined EUVI--COR1--COR2 field of view; the nonradial streamer
path follows the methodology developed by \citet{Alzate2021,Alzate2023}.
We use this established path as the observational framework for a different
physical test.

Events are selected independently in COR2 at $3\,\rsun$ along the frozen
streamer path. The full selection contains 21 outer-coronal anchors, of which
the 12 stronger events form the primary sample. The selection is therefore
defined in the outer corona, without first searching EUVI or COR1 for a
favorable low-coronal signal. Events 9 and 12 were not selected because of
X-/diamond-like morphology; they emerged only after the event-order and
expansion-aware tests were applied to the preselected sample.

Starting from these independently selected COR2 events, we ask whether the
same interval contains an 80--130 minute low-coronal modulation candidate in
EUVI, whether that modulation remains phase-related through COR1, and whether
the later density structures preserve a consistent outward transport pattern.
The candidate release clock, the cross-height phase bridge, and the outward
transfer are therefore tested separately.

The traced path crosses two EUVI height bands and three COR1 height bands
before reaching the COR2-selected structures. EUVI samples the low corona,
where a candidate release modulation may be present, whereas COR1 samples the
streamer toward and above its cusp, where the escaping density pattern may
become more clearly organized.

For the coronagraph analysis we use both total brightness ($\tb$) and
polarized brightness ($\pb$). The SECCHI polarizer sequence provides
co-registered measurements of the same coronal structures in these two
diagnostics. Total brightness maximizes sensitivity to faint moving ridges,
whereas polarized brightness suppresses much of the unpolarized background
and provides a more selective proxy for Thomson-scattered electron column
density. Agreement between $\tb$ and $\pb$ therefore strengthens the
interpretation of a ridge as a real coronal density enhancement rather than
a processing or background feature. Both quantities remain line-of-sight
brightness integrals rather than local plasma measurements.

\subsection{What \texorpdfstring{$pB$}{pB} measures}

White-light coronagraphs observe Thomson scattering by free electrons. Total brightness $\tb$ contains the K-coronal signal together with foreground and instrumental contributions. Polarized brightness $\pb$ isolates the linearly polarized part and is therefore a more selective proxy for electron column density after calibration. It is still a line-of-sight integral, not a local density and not a magnetic-field measurement.

For three polarizer measurements, a simple diagnostic fit is
\begin{equation}
 I(\alpha)=\frac{1}{2}\left[\tb+P_{\rm tan}
 \cos 2(\alpha-\chi)\right],
\label{eq:pol}
\end{equation}
where $\alpha$ is polarizer angle and $\chi$ is the local tangential direction. Agreement of a feature in $\tb$ and $\pb$ makes an unpolarized artifact less likely. It does not supply a compression ratio unless the full calibration and background model are applied.

\subsection{What we call a ridge}

Because a ridge is the elementary observable used in every subsequent test, we
define it before assigning any physical interpretation. At each height we form
a time series along the traced streamer. An outward-moving enhancement produces
a sloping path through a time--radius map. We call this path a \emph{ridge}.

A ridge is therefore a measurement-level feature, not yet a plasma identity:
it may represent a propagating plasma structure, a line-of-sight feature, a
background residual, or an unrelated brightness enhancement. With compatible
outward timing, it can support an advected-PDS interpretation. With a repeatable
fixed location, an upstream--downstream plasma jump, and MHD closure, it may
instead contribute to a shock interpretation. Without these additional tests,
it remains only a ridge.

A matched ridge must appear in both $\tb$ and $\pb$ within a frozen spatial and
temporal tolerance.

Three image representations are used: a temporal band-pass map, a 60-minute base difference, and a radial-gradient-normalized difference. These are correlated views of the same photons. Their agreement tests processing robustness; it is not counted as three independent detections. The projected streamer width is measured from $\pb$ and used to widen the matched filter with height. We denote this width proxy by $A(r)$, while explicitly avoiding the claim that it is the area of one magnetic flux tube.

\begin{figure*}
\centering
\includegraphics[width=0.98\textwidth]{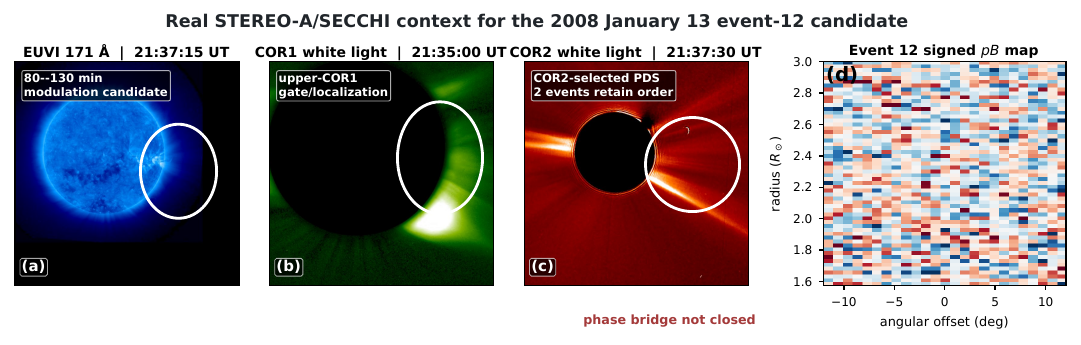}
\caption{Real STEREO-A/SECCHI context for event 12 on 2008 January 13. (a) EUVI 171~\AA\ at 21:37:15 UT, with the low-coronal modulation-candidate sector marked in white. (b) COR1 white light at 21:35:00 UT, marking the upper-COR1 gate/localization region. (c) COR2 white light at 21:37:30 UT, marking the independently selected PDS environment. (d) The corresponding signed 60-minute-base $\pb$ polar map along the traced streamer coordinates. The labels summarize the tested source--gate--transfer picture: two events retain expansion-stable outward order, but a common-delay phase bridge is not detected. Panels (a)--(c) are nearest-time Helioviewer context images; quantitative ridge tests use the frozen quantitative diagnostics, not screen-rendered intensities.}
\label{fig:realcontext}
\end{figure*}

Figure~\ref{fig:realcontext} places event 12 in its real SECCHI EUVI–COR1–COR2 observational context.
The EUVI, COR1, and COR2 panels are nearest-time Helioviewer context views, while the fourth panel is the signed $\pb$ map used by the frozen event analysis. The context views display the nested fields of view and streamer environment and do not enter the reported probabilities.

\section{Experimental Design}\label{sec:methods}

\subsection{Four nested claims}

The analysis proceeds through four claims of increasing physical strength:
\begin{enumerate}
\item \textit{Localization}: COR2-selected events have an unusual ridge response at a particular height.
\item \textit{Event order}: individual events contain several $\tb$--$\pb$ features ordered outward.
\item \textit{Transport}: those features share a physically admissible, common-delay path.
\item \textit{Transition}: a fixed-height plasma jump identifies a stationary MHD branch.
\end{enumerate}
Passing an earlier claim does not imply the next. This hierarchy prevents cell-like morphology from being promoted directly into a shock-diamond claim.

\subsection{Frozen timing and null tests}

For each COR2 anchor, candidate tracks are scanned over a broad coronal speed interval and across the declared image representations. The entire selection is repeated after common shifts of all anchor times. The event-ensemble statistic is the mean best-track score,
\begin{equation}
S_{\rm ens}=\frac{1}{N}\sum_i\max_{v,m}S_i(v,m),
\label{eq:ensemble}
\end{equation}
where $m$ denotes the image representation. The empirical probability is
\begin{equation}
p_{\rm raw}=\frac{N(S_{\rm shift}\geq S_{\rm obs})}{N_{\rm shift}},
\label{eq:praw}
\end{equation}
the fraction of shifted controls whose score equals or exceeds the observed score. Because speed and image representation are reselected within every shifted control, that selection cost is included. Thus $p_{\rm raw}\simeq1$ means that the observed score is common under the shifted-time null, $p_{\rm raw}\simeq0.5$ is unexceptional, and a smaller value means that fewer controls reproduce a score at least as strong. For example, $p_{\rm raw}=0.0279$ means that approximately $2.79\%$ of the shifted controls equal or exceed the real score; it is not the probability that the physical hypothesis is false.

All quoted probabilities are raw empirical values: each describes one declared
diagnostic before any multiple-testing adjustment. The
Benjamini--Hochberg (BH) procedure controls the false discovery rate, that is,
the expected fraction of false positives among results declared significant
when a family of hypotheses is tested simultaneously. Because our diagnostics
address distinct, predeclared physical questions, we report their $\praw$
values separately rather than combine them into one significance measure or
select only the smallest value. Accordingly, no multiplicity-adjusted discovery
claim or population occurrence rate is inferred from a single raw probability.
The horizontal $\praw=0.05$ lines in the figures are therefore visual reference
levels, not discovery boundaries. Conversely, we do not regard a single raw
value below 0.05 as a discovery. A physical chain must pass the connected
predictions in Table~\ref{tab:hypotheses}.

\subsection{Expansion-aware event test}

The two strongest candidates contain several radial nodes at which oppositely
inclined brightness ridges intersect, producing an X-like image-plane
morphology. Such nodes are expected when repeated converging and diverging
fronts form cell-like or diamond-like patterns. Events 9 and 12 are therefore
legitimate shock-cell candidates and motivate the expansion-aware test below;
they are not shock detections. Morphology alone cannot distinguish
underexpanded from overexpanded ordering, which additionally requires the
observed sequence of expansion and compression fronts together with independent
pressure estimates. Establishing a slow, fast, or compound shock branch further
requires a signed plasma jump, upstream and downstream states, and MHD closure.

At each node the filter follows two arms,
\begin{equation}
r_{\pm}(\Delta\theta)=r_v\pm s\Delta\theta,
\label{eq:arms}
\end{equation}
and its width is allowed to be fixed, to grow spherically, or to follow the measured streamer-width proxy:
\begin{equation}
\sigma_r(r)\propto
\left\{1,\ r,\ \sqrt{A(r)/A(r_{\rm in})}\right\}.
\label{eq:width}
\end{equation}
Timing uncertainty is widened from one to two COR1 samples as the front broadens. We then distinguish the probability of simple outward order from the stronger probability that all neighboring nodes lie on one admissible transport chain.

\subsection{What would constitute a shock detection}

A standing-shock interpretation is accepted only if the data show all of the following:
\begin{enumerate}
\item a transition radius that repeats across events or independent folds;
\item an abrupt, signed compression or thermodynamic change in independent diagnostics;
\item a corresponding kink in individually identified event tracks;
\item an upstream speed relative to the transition exceeding the relevant characteristic speed; and
\item downstream consistency with the appropriate Rankine--Hugoniot relations.
\end{enumerate}
COR1 brightness alone cannot supply all five conditions. The test can nevertheless reject a claimed stationary shock when even the necessary morphology and repeatability conditions fail.

\section{Results: From Source-Region Modulation to Event-by-Event Outward Structuring}
\label{sec:results}

\subsection{A localized low-coronal modulation and resonator candidate}

The twelve events were selected independently as outer-coronal COR2 anchors, so the EUVI analysis is a source-side test of their shared interval rather than a signal selected around one favored event. At $1.10$--$1.20\,\rsun$, EUVI 171~\AA\ integrated power in the predeclared 80--130 minute band exceeds all but $3.49\%$ of fitted-AR(1) controls and $2.79\%$ of filter-matched controls ($\praw=0.0349$ and $0.0279$). The largest sampled bin is 128.21 minutes, but the evidence applies to the integrated band, not to a resolved spectral line.

We therefore identify a localized low-coronal modulation and \emph{resonator candidate}, not an MHD eigenmode. Mode identification would require independent calibration and cross-channel phase, velocity, or displacement information. More importantly for the present experiment, a propagation-delayed phase bridge from EUVI to the later outer-coronal events is not recovered. The low-coronal timescale and the later PDS-like organization coexist along the same path, but the modulation is not established as the trigger of any individual release.

Across the observing chain, three results are independently positive: the local 80--130 minute EUVI excess, the population-level upper-COR1 organization, and expansion-stable outward order in two events. The fixed-continuation phase tests do not connect those levels into one rigid train: the EUVI--COR1 test gives $\praw=0.346$, and the EUVI--COR2 continuation gives $\praw=1$.

\begin{figure*}[!t]
\centering
\includegraphics[width=0.98\textwidth]
{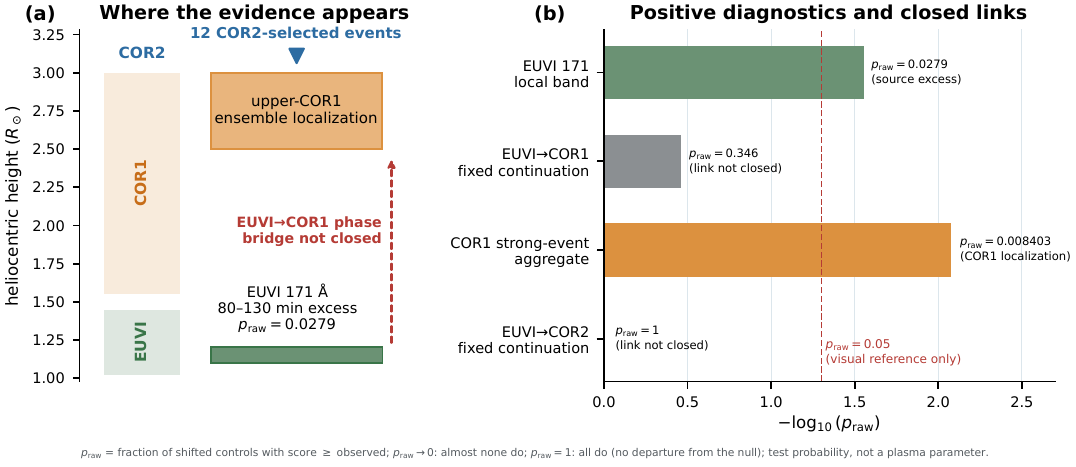}
\caption{Where the possible low-coronal PDS clock appears---and where its phase
connection is lost. Here \emph{clock} means a recurring modulation that may set
the cadence of PDS release, not the PDS structures themselves. We test whether
the twelve outward-moving COR2 density structures can be
traced back to a repeating low-coronal modulation. Panel (a) shows a localized
80--130 minute EUVI 171~\AA\ modulation candidate near $1.1\,\rsun$ and the
strongest common organization of the events in upper COR1 at
$2.5$--$3.0\,\rsun$. The red arrow shows that a phase-preserving EUVI--COR1
connection is not recovered. Panel (b) summarizes four empirical diagnostics;
smaller $p_{\rm raw}$ means that shifted controls reproduce the observed
pattern less often. The EUVI modulation and COR1 localization are positive
diagnostics, but neither the EUVI--COR1 nor EUVI--COR2 phase link closes. The
data therefore support a localized low-coronal modulator or resonator candidate
that may contribute to later PDS structuring, but not one continuously
phase-locked PDS train from the low corona into COR2.}
\label{fig:reschain}
\end{figure*}

Figure~\ref{fig:reschain} summarizes this hierarchy of evidence: the local EUVI and ensemble COR1 diagnostics are positive, whereas the tested cross-height phase links remain open. The following subsections examine the upper-COR1 localization and the event-resolved outward ordering separately.

\subsection{The strongest ensemble localization occurs in upper COR1}

\begin{figure*}[!t]
\centering
\includegraphics[width=0.98\textwidth]{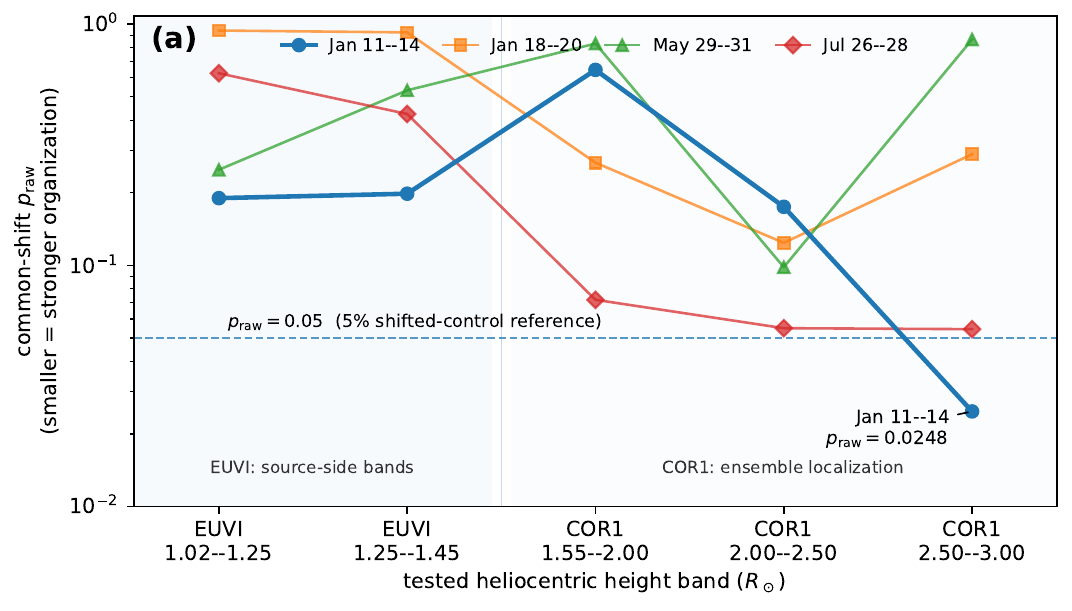}
\caption{Height localization of the 12-event ensemble. The minimum raw common-shift probability occurs in upper COR1 at $2.5$--$3.0\,\rsun$ ($\praw=0.0248$), marking the strongest ensemble organization. Lines connect tests across height bands and are not propagation trajectories.}
\label{fig:height}
\end{figure*}

The strongest height-localization result occurs in upper COR1, $2.5$--$3.0\,\rsun$, where the common-shift probability reaches $\praw=0.0248$ (Figure~\ref{fig:height}). This statistic measures how rarely shifted control ensembles reproduce an association as strong as the observed one; it is a localization result, not a brightness amplitude or a claimed discovery threshold. The clearest common organization therefore appears near the cusp/current-sheet transfer domain rather than in the tested EUVI bands.

This localization does not identify the ultimate source or prove a stationary transition. It instead marks the height range in which the COR2-selected population becomes most coherently organized, making upper COR1 the natural region in which to test gate dynamics, outward transport, and possible shock-cell processing.

\subsection{The 12 events do not preserve the same outward structure}

Table~\ref{tab:eventphysics} gives the physical reading of every primary event. Two events retain outward order after expansion is allowed (Class A). Four contain several compatible coronal ridges but do not close timing (Class B). Two are weak multi-ridge cases (Class C). Four contain too few connected features to define a train (Class D).

\begin{table*}[!t]
\centering
\caption{Physical reading of the 12 independently selected COR2 events. The table is deliberately descriptive: it shows how the sample narrows from a mixed PDS population to the two events that retain expansion-stable outward order and then motivate the shock-cell test.}
\label{tab:eventphysics}

{\fontsize{7.55}{8.45}\selectfont
\setlength{\tabcolsep}{2.7pt}
\renewcommand{\arraystretch}{1.08}
\begin{tabular}{llll}
\toprule
\parbox[t]{0.055\textwidth}{Event} & \parbox[t]{0.23\textwidth}{Observed density pattern} & \parbox[t]{0.055\textwidth}{Class} & \parbox[t]{0.57\textwidth}{Physical reading} \\
\midrule
1 & \parbox[t]{0.23\textwidth}{4 matched ridges} & B & \parbox[t]{0.57\textwidth}{Strong multi-ridge morphology, but the timing does not define one unique source-to-COR2 history.} \\
2 & \parbox[t]{0.23\textwidth}{2 ridges} & D & \parbox[t]{0.57\textwidth}{Partial sequence; too short to define a PDS train.} \\
3 & \parbox[t]{0.23\textwidth}{1 isolated ridge} & D & \parbox[t]{0.57\textwidth}{Localized enhancement; outward transport is not resolved.} \\
4 & \parbox[t]{0.23\textwidth}{3 quasi-regular ridges} & B & \parbox[t]{0.57\textwidth}{Repeated multi-ridge structure with a substantial $\tb$--$\pb$ timing offset.} \\
5 & \parbox[t]{0.23\textwidth}{1 isolated ridge} & D & \parbox[t]{0.57\textwidth}{Timing hint without a resolved outward sequence.} \\
6 & \parbox[t]{0.23\textwidth}{3 weak ridges} & C & \parbox[t]{0.57\textwidth}{Weak multi-ridge structure with strongly evolving $\tb$--$\pb$ timing.} \\
7 & \parbox[t]{0.23\textwidth}{1 isolated ridge} & D & \parbox[t]{0.57\textwidth}{One ridge; the formal path fit is insufficient to define a train.} \\
8 & \parbox[t]{0.23\textwidth}{3 irregular ridges} & C & \parbox[t]{0.57\textwidth}{Compatible geometry, but only weak outward ordering.} \\
\textbf{9} & \parbox[t]{0.23\textwidth}{\textbf{4 ordered nodes}} & \textbf{A} & \parbox[t]{0.57\textwidth}{\textbf{Expansion-stable outward order; four-node X-/cell-like morphology.}} \\
10 & \parbox[t]{0.23\textwidth}{3 quasi-regular ridges} & B & \parbox[t]{0.57\textwidth}{Regular morphology with an evolving phase relation.} \\
11 & \parbox[t]{0.23\textwidth}{3 ridges} & B & \parbox[t]{0.57\textwidth}{Multi-ridge pattern compatible with outward propagation, without common-ridge closure.} \\
\textbf{12} & \parbox[t]{0.23\textwidth}{\textbf{3 ordered nodes}} & \textbf{A} & \parbox[t]{0.57\textwidth}{\textbf{Expansion-stable outward order; three-node X-/cell-like morphology, with the outer node in the upper-COR1 organization domain.}} \\
\bottomrule
\end{tabular}
}

\begin{minipage}{0.97\textwidth}
\vspace{2pt}\footnotesize
\textit{Note.} Class A is the positive event-order result; B denotes substantial multi-ridge structure without unique timing closure; C weak/irregular multi-ridge structure; and D an isolated or incomplete sequence. The classes organize the morphology and transport evidence only. They do not identify an MHD mode or shock branch.
\end{minipage}
\end{table*}

\paragraph{Numerical audit.}
The complete frozen event statistics are retained in Table~\ref{tab:events}. The fitted accelerating or decelerating family is an alignment preference within the tested grid, not a direct measurement of the bulk solar-wind acceleration.

\begin{table*}[!t]
\centering
\caption{Frozen numerical audit underlying the event classes. The three raw empirical probabilities are shown together as $p_{\rm cell}/p_{\rm best}/p_{\rm common}$ so the table remains readable at journal width; events 9 and 12 are highlighted because they alone retain expansion-stable outward order.}
\label{tab:events}

{\fontsize{7.65}{8.5}\selectfont
\setlength{\tabcolsep}{4.0pt}
\renewcommand{\arraystretch}{1.06}
\begin{tabular}{c l c c c l c}
\toprule
Event & UTC & $N_{\rm pk}$ & $p_{\rm cell}/p_{\rm best}/p_{\rm common}$ & $|\Delta t_{tB-pB}|$ (min) & Best path fit (km s$^{-1}$) & Class \\
\midrule
1 & 01-11 14:37 & 4 & 0.069 / 0.743 / 0.042 & 30 & dec. 300$\to$50 & B \\
2 & 01-12 05:07 & 2 & 0.618 / 0.854 / 0.312 & 75 & dec. 300$\to$20 & D \\
3 & 01-12 11:37 & 1 & 0.944 / 0.431 / 0.125 & 75 & dec. 250$\to$20 & D \\
4 & 01-12 22:07 & 3 & 0.076 / 0.118 / 0.479 & 60 & dec. 200$\to$25 & B \\
5 & 01-13 01:07 & 1 & 0.944 / 0.438 / 0.069 & 90 & dec. 300$\to$20 & D \\
6 & 01-13 05:37 & 3 & 0.146 / 0.938 / 0.688 & 180 & const. 300 & C \\
7 & 01-13 10:07 & 1 & 0.944 / 0.722 / 0.375 & 45 & acc. 30$\to$250 & D \\
8 & 01-13 14:37 & 3 & 0.181 / 0.646 / 0.729 & 15 & dec. 100$\to$40 & C \\
\textbf{9} & 01-13 15:37 & \textbf{4} & \textbf{0.028 / 0.486 / 0.257} & \textbf{30} & \textbf{dec. 100$\to$30} & \textbf{A} \\
10 & 01-13 17:07 & 3 & 0.090 / 0.542 / 0.236 & 135 & dec. 150$\to$20 & B \\
11 & 01-13 19:37 & 3 & 0.167 / 0.062 / 0.479 & 30 & acc. 80$\to$300 & B \\
\textbf{12} & 01-13 21:37 & \textbf{3} & \textbf{0.104 / 0.035 / 0.833} & \textbf{0} & \textbf{dec. 100$\to$25} & \textbf{A} \\
\bottomrule
\end{tabular}
}

\begin{minipage}{0.97\textwidth}
\vspace{2pt}\footnotesize
\textit{Note.} Smaller raw $p$ means that fewer shifted controls reproduce a score at least as strong as the observed one; the three diagnostics answer different questions and are not multiplied into one significance. ``acc.'', ``dec.'', and ``const.'' denote the trajectory family that best aligns the ridges within the tested grid, not a direct measurement of solar-wind acceleration. The $\tb$--$\pb$ offset is a timing diagnostic. Class A/B/C/D has the physical meaning given in Table~\ref{tab:eventphysics}.
\end{minipage}
\end{table*}

\begin{figure*}
\centering
\includegraphics[width=0.98\textwidth]{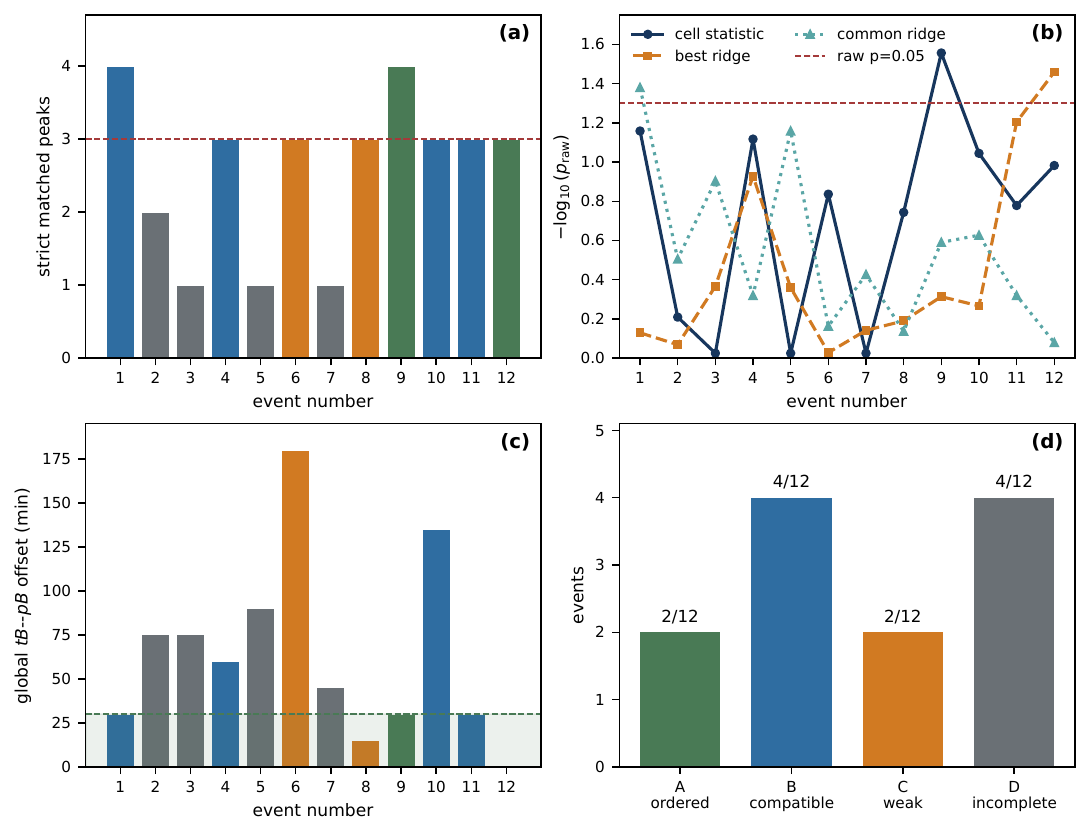}
\caption{The 12-event survey. (a) Number of strict $\tb$--$\pb$ matched peaks. (b) Raw empirical diagnostics. (c) Global timing difference between $\tb$ and $\pb$. (d) Physical classes. Events 9 and 12 are the two expansion-stable ordered candidates; four cases never contain enough connected peaks to define a train.}
\label{fig:survey}
\end{figure*}

The two event tables play different roles: Table~\ref{tab:eventphysics} tells the physical story, while Table~\ref{tab:events} preserves the numerical audit. Figure~\ref{fig:survey} then shows visually why the classification cannot be replaced by a single average. Event 1 has four matched peaks and the strongest common morphology outside the two pilot events. Event 4 has a nearly regular three-cell pattern, but $\tb$ and $\pb$ disagree in time. Event 10 is morphologically regular and phase-inconsistent. Event 11 is compatible with an accelerating pattern but has no common-ridge closure. Events 9 and 12 are the only cases that combine several nodes with expansion-stable outward order.

The event diversity is part of the physics. If one universal resonator or one stationary shock controlled every event, the sample should not divide so sharply into ordered, compatible, weak, and incomplete cases under the same observing geometry. This event-to-event diversity instead points toward intermittent source--gate coupling. Strict phase locking is not recovered, but the underlying density structuring can persist while its timing evolves during outward propagation.

\subsection{The two ordered events survive an expanding streamer}

\begin{figure*}
\centering
\includegraphics[width=0.98\textwidth]{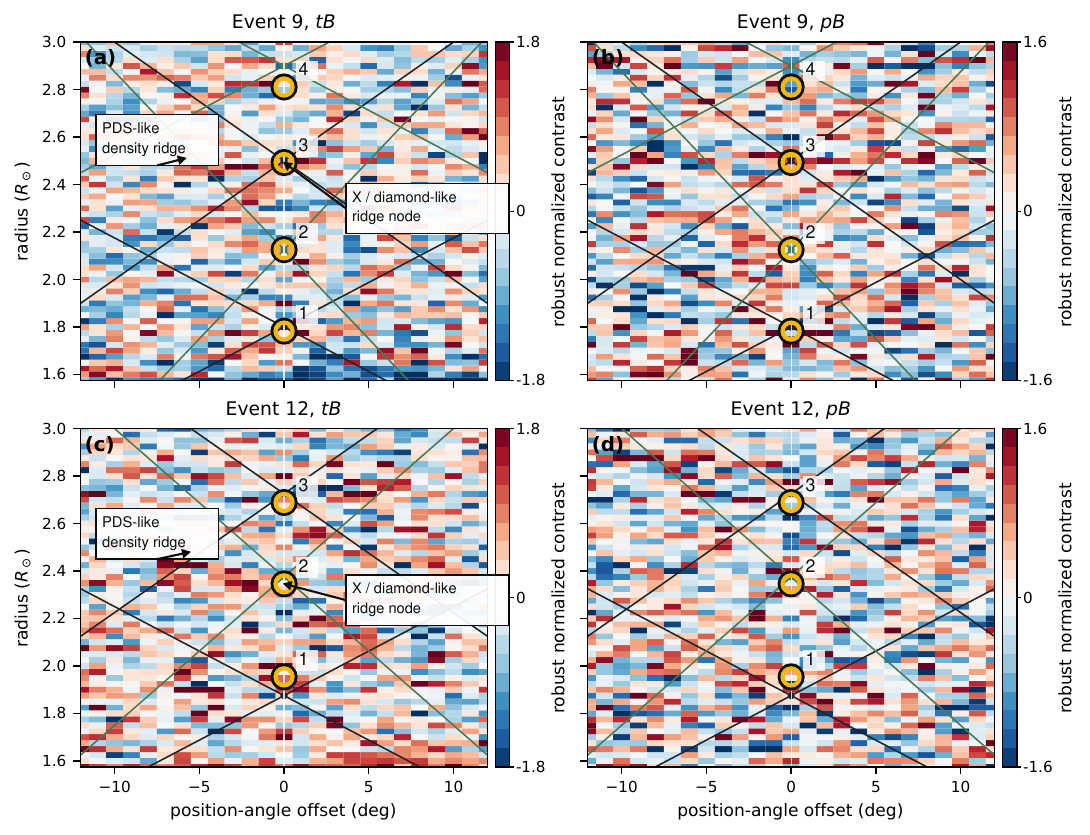}
\caption{Polar maps of events 9 and 12 in total and polarized brightness. The frozen expansion-aware matched-filter arms identify four nodes in event 9 at $1.775$, $2.125$, $2.500$, and $2.825\,\rsun$, and three in event 12 at $1.950$, $2.350$, and $2.700\,\rsun$. The annotations point to representative PDS-like density ridges and to the numbered X-/diamond-like ridge intersections; they are visual guides to the same frozen measurements, not additional features selected by eye. The outer nodes of both events enter the independently identified upper-COR1 organization domain. The maps establish structured, outward-ordered shock-cell morphology; stationarity and MHD-shock classification require the stronger plasma-transition test.}
\label{fig:maps}
\end{figure*}

Events 9 and 12 are the only two cases that retain expansion-stable outward ordering of several individually resolved structures. Their frozen image-plane node radii are
\begin{align*}
\text{event 9:}&\quad 1.775,\ 2.125,\ 2.500,\ 2.825\,\rsun,\\
\text{event 12:}&\quad 1.950,\ 2.350,\ 2.700\,\rsun.
\end{align*}
Thus the outer nodes of both events fall inside the independently identified $2.5$--$3.0\,\rsun$ upper-COR1 organization domain. Figure~\ref{fig:maps} shows the corresponding $\tb$ and $\pb$ polar maps. The oblique supports are the frozen expansion-aware matched-filter arms; their intersections are observed image-plane vertices, not automatically deprojected shock locations. A compact visual guide to the two distinct measurements---the time--radius definition of a candidate PDS train and the polar-map definition of X-/diamond-like ridge nodes---is given in Appendix~\ref{app:measurementguide} (Figures~\ref{fig:app_pds} and~\ref{fig:app_xnodes}).

\begin{figure*}
\centering
\includegraphics[width=0.96\textwidth]{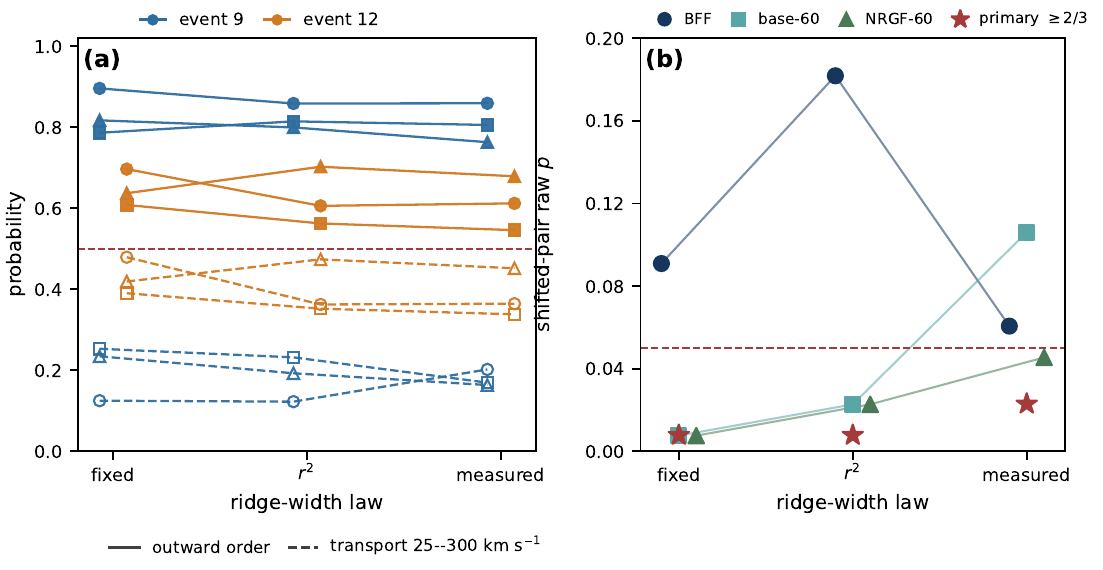}
\caption{Expansion-aware test. In panel (a), solid curves give the probability of outward node order and dashed curves the stronger probability that all neighboring-node speeds remain within $25$--$300\,\kms$. Order remains robust for events 9 and 12 under fixed, spherical, and measured-width filters, while the full transport probability remains below 0.5 for both events. Panel (b) compares the real pair with 131 fixed-separation shifted controls in BFF, base-60, and NRGF-60 representations. The primary $\geq2/3$ raw probabilities are 0.0076, 0.0076, and 0.0227 for the three width laws, respectively. The result therefore supports expansion-stable order, not one closed common-delay transport chain.}
\label{fig:expansion}
\end{figure*}

Outward order survives all three streamer-width laws in Figure~\ref{fig:expansion}, so the result is not an artifact of treating the corona as a rigid tube. The stronger common-delay condition is not recovered: the node-to-node delays are too broad or event-dependent to define one unique transported pulse train. The shifted-pair control nevertheless remains unusual after the measured-width correction, supporting the reality of organized morphology in the pair.

\subsection{Shock-cell morphology is present; MHD classification remains open}

Events 9 and 12 show the geometric ingredients sought in the shock-cell test: oppositely inclined fronts, repeated X-like ridge intersections, cell-like spacing, and expansion-stable outward order. We therefore report \emph{shock-cell morphology} in these two events. The node radii are not identical between the pair, so the data do not establish one universal stationary transition height; this does not by itself exclude quasi-stationary shock-cell processing within each event.

The available brightness data stop short of a full stationary-shock closure. Limited $\pb$ compression candidates accompany the morphology, but they are not reproduced as a diagnostic-independent upstream/downstream plasma jump; for event 12 the corresponding $\tb$ jump is unavailable. The best nonlinear path also does not improve predictively over the constant-speed alternative, and the nominal event-12 phase kink is not significant. Establishing a stationary MHD shock and classifying a slow, fast, or compound branch would additionally require independent plasma states, magnetic field and temperature constraints, the normal flow/shock speed, characteristic speeds, and Rankine--Hugoniot consistency.

The result is therefore two-tiered: the shock-cell pattern is an observed morphological result, while stationary-shock identification and slow/fast/compound MHD classification remain open rather than excluded. In a stationary shock-cell train the oblique compression/shock fronts and their vertices can remain approximately fixed in the flow frame; the measured nodes are therefore plausible stationary-vertex candidates, not individual shock detections. Demonstrating stationarity requires repeated localization at the same radii in independent times or folds together with the plasma-transition diagnostics above.

\subsection{The calibration control sets the present evidential limit}

A direct reconstruction from COR1 Level-0 polarizer triplets passes the image-quality screen but does not reproduce the favorable aggregate association or a stable acceleration/deceleration break. This disagreement is therefore retained as an unresolved calibration/background systematic rather than interpreted as a failed physical test. Level-0 data are not intrinsically less reliable for morphology: in some cases they preserve native image structure and avoid processing-dependent modifications introduced at later stages. The favorable reduction therefore remains a valid diagnostic result, while an independent reprocessing test with an alternative calibration/background pathway would provide an additional robustness check.

\section{Physical Interpretation and a New PDS Diagnostic}\label{sec:discussion}

\subsection{Source modulation and outward order are distinct observables}

The experiment does not recover one coherent 80--130 minute phase from EUVI through COR1 and COR2, so it does not support one continuously visible clock for all twelve events. The low-coronal excess nevertheless remains a positive local result: it identifies a band-limited modulation candidate. The phase test is essential because period coexistence alone cannot tell whether that modulation controls the release cadence. A compact MHD response may modulate the probability of reconnection when the cusp/current sheet is receptive, while intrinsic tearing provides an alternative source of quasi-periodicity.

The event-to-event diversity adds an independent constraint. Two events retain expansion-stable order, four show compatible multi-ridge structure, two are weak, and four are incomplete. Such a mixed population is more naturally accommodated by intermittent coupling than by one identical release-and-transfer history for every event. In this sense, failure of a universal phase relation does not erase the source-side signal; it limits how directly that signal can be mapped onto individual later PDS.

This is a physical result rather than only a statistical non-detection. If all
twelve disturbances were passive copies released by one stable clock and
carried through one stationary transfer function, a common propagation delay
should align their phases after allowance for acceleration and measurement
uncertainty. Instead, spatial order survives in two events while exact phase
does not survive across the ensemble. The data therefore distinguish
\emph{structural memory} from \emph{phase memory}: the expanding corona can
preserve which density enhancement lies ahead of another even when an
intermittent gate, changing propagation speed, or evolving geometry destroys a
single clock-to-event mapping. Near-regular spatial separation is therefore not equivalent to phase locking: event-dependent delays can shift the arrival phase while leaving the radial sequence intact, provided neighboring structures do not overtake one another (Appendix~\ref{app:measurementguide}).

\subsection{Distinct heights, one connected physical chain}

The positive signatures are not co-spatial, and that separation is physically informative. The low-coronal modulation candidate is detected in EUVI 171~\AA\ at $1.10$--$1.20\,\rsun$, whereas the 12-event ensemble becomes most strongly organized much higher, in upper COR1 at $2.5$--$3.0\,\rsun$. The ordered structures occupy the intervening and upper-coronal range: event 9 has nodes at $1.775$, $2.125$, $2.500$, and $2.825\,\rsun$, while event 12 has nodes at $1.950$, $2.350$, and $2.700\,\rsun$. Thus the low-coronal modulation and the shock-cell morphology are not two measurements of one structure at one height. Instead, the outer nodes of both ordered events enter the independently identified upper-COR1 organization domain.

This ordering suggests a source--processing sequence rather than a spatial coincidence. A low-coronal MHD response or intrinsic reconnection can supply temporal variability; a cusp/current-sheet gate can select which perturbations become releases and introduce event-dependent delay; and the outward flow can then preserve radial order while exact phase evolves. The missing phase bridge is therefore not evidence that the three positive results are unrelated. It rules out the stronger picture in which one continuously visible, phase-locked object is followed from the low corona to COR2. The fact that the twelve events were selected independently in COR2 before the low-coronal and shock-cell tests further separates this interpretation from a morphology-selected construction.

The shock-cell result is naturally placed at the transfer/processing stage of this chain. Events 9 and 12 show oppositely inclined fronts, repeated X-like intersections, cell-like spacing, and expansion-stable order. Their outer nodes overlap the height range where the ensemble organization is strongest, making stationary or quasi-stationary shock-cell processing a physically viable interpretation. The node radii differ between the two events, so the data do not support one universal fixed shock height. That difference does not, however, exclude approximately stationary shock-cell vertices within an individual event: in a stationary shock-cell train, oblique compression/shock fronts and their vertices can remain approximately fixed in the flow frame.

The observational claim therefore has two levels. Shock-cell morphology is present in the two strongest ordered events and is a positive result. Demonstrating that the nodes are stationary MHD shocks is a stronger claim that requires repeated localization of a transition radius together with independent upstream and downstream plasma states, magnetic-field and temperature constraints, normal flow and shock speeds, characteristic speeds, and Rankine--Hugoniot consistency. Those measurements are also required to classify a slow, fast, or compound branch. The present data leave that classification open rather than excluding shock processing.

The overlap between the upper-COR1 ensemble maximum and the outer nodes of
events 9 and 12 is especially suggestive. It connects two measurements chosen
in different ways: a population-level localization from all twelve
COR2-selected events and event-resolved X-/cell-like geometry in the only two
cases whose outward order survives expansion. This conjunction has not been
used to select the events and is therefore a new observational clue. It does
not prove that shocks create the coherence, because the sample is small and
the present brightness data lack plasma closure. It does identify a sharply
testable possibility: upper COR1 may be the region where an initially weak or
intermittent density sequence becomes organized or reprocessed as it enters
the accelerating wind.

\subsection{What is new and why it matters for the young solar wind}

Earlier SECCHI work established that disturbances can be traced along this
nonradial path from the low to the high corona \citep{Alzate2024}. The present
experiment adds four new constraints. First, it isolates a local 80--130 minute
EUVI modulation candidate without promoting the 128.21 minute maximum bin to a
resolved spectral line. Second, it locates the strongest ensemble organization
in upper COR1 rather than assuming that the first visible low-coronal signal is
also the point where the PDS train becomes coherent. Third, it shows that the
twelve-event population is physically mixed rather than one phase-locked
family. Fourth, it identifies expansion-stable outward order and shock-cell
morphology in events 9 and 12, whose outer nodes enter the independently found
upper-COR1 organization domain.

These results turn PDS into diagnostics of the acceleration region rather than
merely periodicities to be cataloged at large distance. Cadence tests a possible
source clock; cross-height phase tests causal continuity; radial order tests
whether the transfer preserves the sequence; width and separation test
expansion and acceleration; $\tb$--$\pb$ morphology tests whether the same
density-bearing structures are recovered; and fixed nodes plus a plasma jump
would test stationary MHD processing. No single diagnostic is sufficient, but
their combination separates source, gate, and transport physics.

The timing is favorable for such an approach. High-cadence EUV observations,
coronagraph polarimetry, multi-viewpoint imaging, and inner-heliospheric in-situ
measurements can now be compared with forward-modeled observables rather than
with idealized density fields alone. The practical challenge for observers and
data analysts is to retain calibrated signed-polarization information and
event-level uncertainties; for numerical modelers it is to generate synthetic
$\tb$ and $\pb$ images from time-dependent streamer/cusp calculations. A model
that reproduces a period but not the observed height localization, phase
jitter, expansion-stable node order, and polarization morphology would not
reproduce the experiment.

The result is therefore intended as a trigger for a broader program rather
than a closed mechanism claim. Larger samples can test whether upper-COR1
organization is common, whether it shifts with streamer geometry or solar-wind
speed, and whether shock-cell candidates preferentially occur in the most
organized events. Multi-viewpoint tracking can determine whether the apparent
nodes are finite three-dimensional vertices. Spectroscopy, density inversion,
and characteristic-speed estimates can then decide whether the observed cells
are slow, fast, compound, or non-shock structures. Each failure mode is
informative because it removes one link from the source--gate--transfer chain.

\subsection{Synthesis}

\begin{figure*}[!t]
\centering
\includegraphics[width=0.80\textwidth]{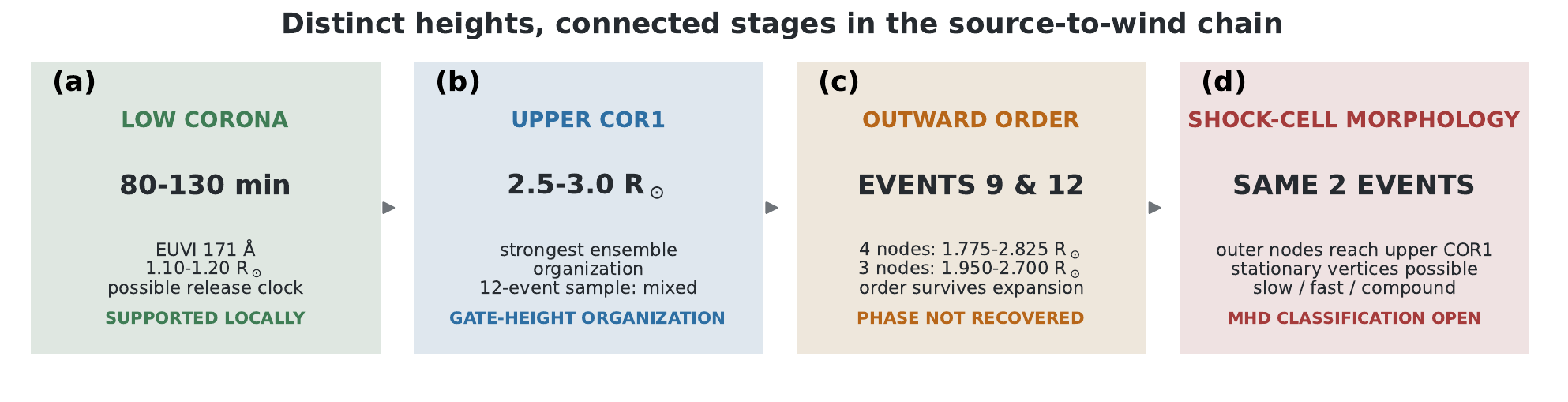}
\caption{Final synthesis: a local clock candidate, strongest upper-COR1 organization, expansion-stable order in events 9 and 12, and shock-cell morphology with open MHD classification.}
\label{fig:synthesis}
\end{figure*}

The experiment separates three observational stages rather than forcing them to coincide: a possible release clock low in the corona, preferred organization near the upper-COR1 gate/transfer domain, and shock-cell morphology in the two events whose outward order survives expansion. The positive result is therefore an ordered set of constraints, not a closed causal chain. The independently selected COR2 population becomes most strongly organized in upper COR1, and the outer nodes of events 9 and 12 enter that same domain. Spatial order is more robust than exact phase, favoring intermittent source--gate--transfer coupling over one universally phase-locked PDS train. Figure~\ref{fig:synthesis} summarizes these distinct stages and open links.

The open links refine rather than erase this picture. A unique common-delay phase bridge is not recovered, so the low-coronal modulation is not established as the trigger of individual later releases. Likewise, the observed shock-cell morphology and limited compression signatures do not yet provide the independent upstream/downstream plasma-state and characteristic-speed closure needed for a stationary MHD-shock identification. Thus the data support neither one universal clock nor one universal fixed shock height, while remaining compatible with a low-coronal modulator and with stationary or quasi-stationary shock-cell processing in individual events.

Together these results favor a dynamic source--gate--transfer picture in which intrinsic tearing or a compact oscillatory perturbation supplies variability, a cusp/current-sheet gate selects and delays releases, and the expanding flow preserves radial order while timing can evolve. The natural next calculation is an expanding, time-dependent cusp/current-sheet model with synthetic $\tb$ and $\pb$ observables, tested against the measured phase, width, node, and plasma-transition diagnostics.

\section{Falsifiable Next Experiment}\label{sec:next}

A stronger independent confirmation should repeat the analysis with an alternative SECCHI calibration/background pathway and require:
\begin{enumerate}
\item a source-side signal in two independent low-coronal observables across the full 80--130 minute band;
\item event-resolved tracking through adjacent heights with expansion-aware widths and 15--30 minute timing uncertainty;
\item at least three outward nodes in both $\tb$ and signed $\pb$;
\item multi-viewpoint geometry for any finite vertex or fan claim;
\item a repeatable jump radius with background controls; and
\item an independent magnetosonic-speed estimate before assigning a slow, fast, or compound branch.
\end{enumerate}
Phase-locked low-coronal modulation followed by repeated release would favor a resonator-controlled gate; plasmoid trains without upstream phase coherence would favor intrinsic tearing; a fixed-height jump with MHD closure would establish stationary shock processing; and growing phase jitter with changing cusp geometry would support a reforming gate.

\section{Conclusions}\label{sec:conclusion}

We used an independently selected outer-coronal PDS sample to ask where observable organization appears and how much of a possible low-coronal cadence survives through the coronal transfer region.

\begin{enumerate}
\item A calibrated EUVI 171~\AA\ diagnostic shows a local 80--130 minute excess at $1.10$--$1.20\,\rsun$ ($\praw=0.0279$). We identify a low-coronal modulation/resonator candidate, not an MHD mode or an event-by-event trigger; the cross-height phase bridge and independent calibration confirmation remain open.
\item The 12 COR2-selected events are most strongly organized in upper COR1 at $2.5$--$3.0\,\rsun$ ($\praw=0.0248$), localizing the clearest common organization near the cusp/current-sheet transfer domain.
\item The sample shows clear event-to-event diversity: two events are expansion-stable and ordered, four are compatible multi-ridge cases, two are weak, and four are incomplete. Thus one universal phase-locked transfer history is not required by the data.
\item Events 9 and 12 contain four and three ordered nodes spanning $1.775$--$2.825\,\rsun$ and $1.950$--$2.700\,\rsun$. Their outer nodes enter the upper-COR1 organization domain, and their X-/cell-like geometry constitutes the strongest shock-cell morphology in the sample. The nodes could be approximately stationary vertices of a shock-cell train, but the present data do not establish stationarity; MHD-shock identification and slow/fast/compound classification require additional plasma-state and characteristic-speed closure.
\item Taken together, the observations favor an intermittent source--gate--transfer picture: a low-coronal modulator or intrinsic reconnection supplies variability, a cusp/current-sheet gate selects and processes releases, and the expanding wind advects the resulting density structures while exact phase can evolve.
\end{enumerate}

The deepest result is therefore methodological and physical at once. A PDS period alone cannot identify its origin. Source modulation, phase, event order, differential delay, front width, shock-cell morphology, and plasma-transition diagnostics together make PDS probes of how the corona opens into the solar wind.

\appendix
\section{Measurement guide: how candidate PDS trains and X-/diamond-like nodes are identified}
\label{app:measurementguide}

\begin{figure*}[!t]
\centering
\includegraphics[width=0.98\textwidth]{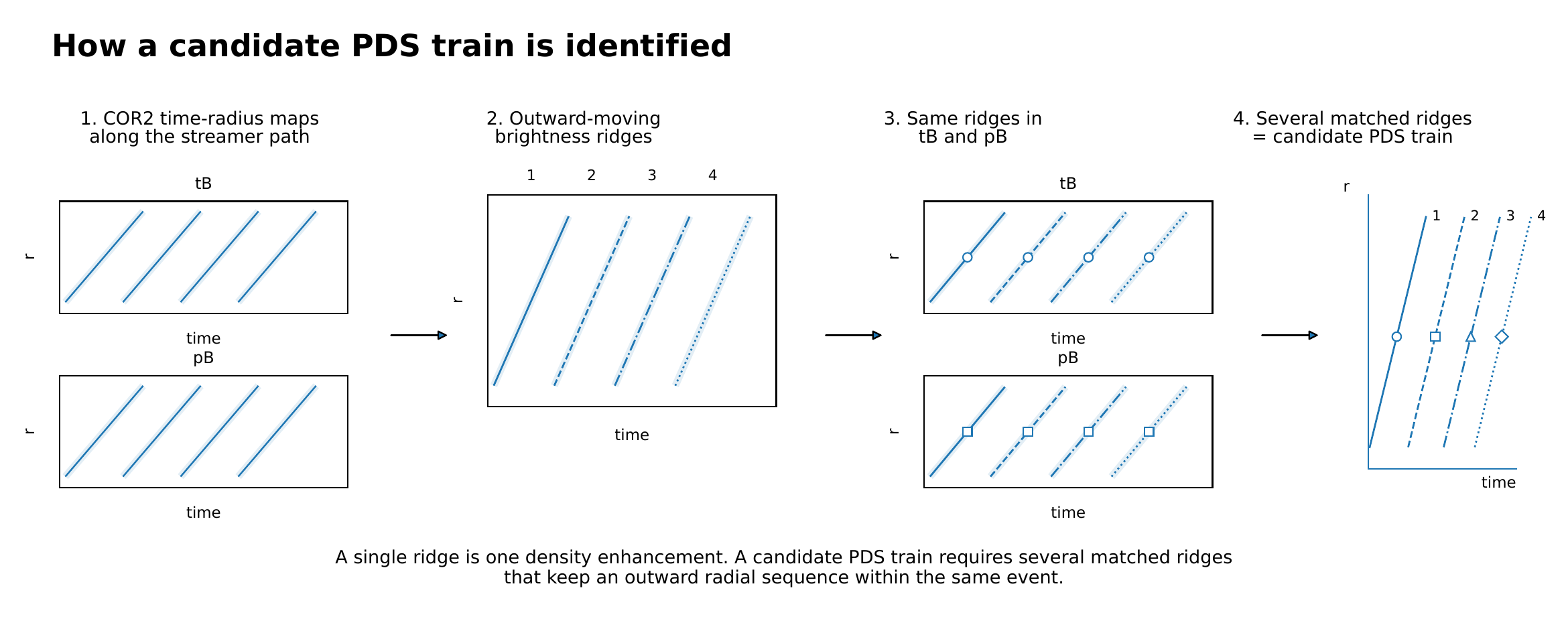}
\caption{Schematic guide to the PDS measurement. COR2 time--radius maps are sampled along the traced streamer path in $\tb$ and $\pb$. Outward-moving brightness ridges are identified, corresponding ridges are matched between the two diagnostics, and several matched ridges in one event define a candidate PDS density train. A single ridge is one density enhancement, not a train. The schematic illustrates the measurement logic only; all event statistics in the main text use the frozen analysis data and the declared matched-filter procedure.}
\label{fig:app_pds}
\end{figure*}

\begin{figure*}[!t]
\centering
\includegraphics[width=0.98\textwidth]{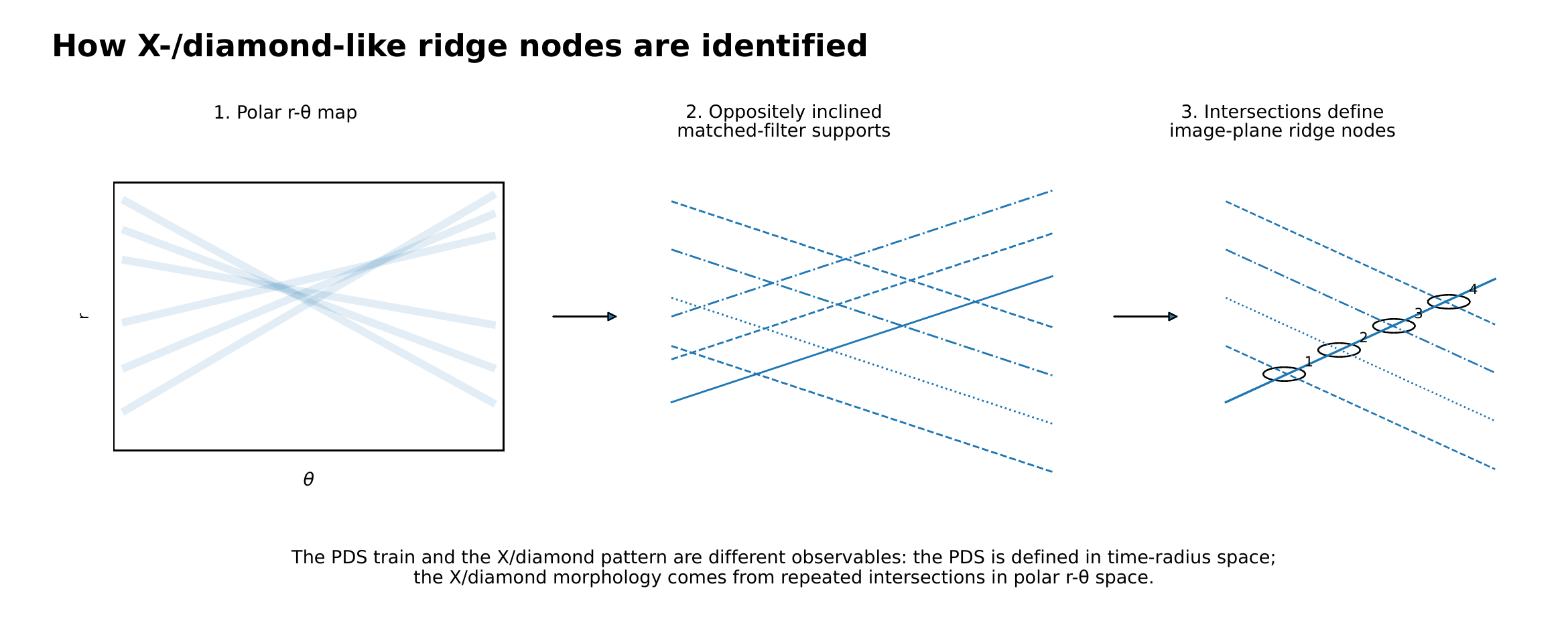}
\caption{Schematic guide to the X-/diamond-node measurement. In a polar $r$--$\theta$ map, oppositely inclined support ridges returned by the frozen matched filter intersect at measured image-plane nodes. A repeated sequence of such intersections with cell-like spacing produces the X-/diamond-like morphology reported for events 9 and 12. This geometrical observable is distinct from the time--radius definition of the PDS train in Figure~\ref{fig:app_pds}.}
\label{fig:app_xnodes}
\end{figure*}

The two observational patterns used in the event analysis are related, but they are not the same measurement. A candidate PDS train is defined from the temporal and radial organization of density enhancements in COR2, whereas X-/diamond-like morphology is defined from the geometry of ridge intersections in a polar $r$--$\theta$ representation. Figures~\ref{fig:app_pds} and~\ref{fig:app_xnodes} give a schematic guide to these two steps so that the measurements in Figures~\ref{fig:survey}--\ref{fig:maps} can be read directly from the data products.

\subsection{How a candidate PDS train is identified}

The PDS identification starts from COR2 time--radius maps sampled along the frozen streamer path. In a single brightness diagnostic, one outward-moving ridge represents one density enhancement; a single ridge is therefore not a PDS train. Candidate ridges are compared between total brightness ($\tb$) and polarized brightness ($\pb$), and several matched ridges within the same event are required before the structure is described as a candidate PDS density train. The additional event-order test asks whether those density enhancements retain the same outward radial sequence after propagation and streamer expansion are allowed. Thus the defining PDS observable is a sequence in time--radius space, not the presence of an X-shaped pattern.

\subsection{How X-/diamond-like nodes are identified}

The X-/diamond-like test is performed separately in polar $r$--$\theta$ maps for the strongest outward-order candidates. The frozen expansion-aware matched filter returns ridge-support arms with opposite inclinations. Their intersections define image-plane ridge nodes; repeated intersections with cell-like spacing form the observed X-/diamond-like morphology. These nodes are geometrical measurements in the image plane. They are not automatically deprojected nozzle vertices and they are not individual shock detections. Promoting the morphology to a stationary slow, fast, or compound MHD-shock identification requires the stronger plasma-transition and characteristic-speed closure described in Section~\ref{sec:next}.

\subsection{Why outward order can survive when exact phase does not}

Phase and order encode different information. Phase measures the timing of a later structure relative to the low-coronal 80--130 minute modulation. If successive structures acquire event-dependent delays $\tau_n$ in a reforming release gate or in the propagation path, their arrival intervals satisfy
\begin{equation}
P_n(r)=P_n^{(0)}+\tau_{n+1}(r)-\tau_n(r).
\end{equation}
The additional delay term can shift the phase and make the intervals only approximately regular. The spatial sequence can nevertheless remain $A\!\rightarrow\!B\!\rightarrow\!C\!\rightarrow\!D$ as long as the differential delays are not large enough for neighboring structures to overtake one another. Therefore near-regular ridge or node spacing is not evidence by itself for phase locking to the low-coronal clock. This is the observational distinction used in the main text: events 9 and 12 retain expansion-stable outward order, whereas one unique EUVI--COR1--COR2 common-delay phase sequence is not recovered.

\section*{Data and Software Availability}

The analysis code and compact frozen analysis results associated with this study are publicly archived on Zenodo at \url{https://doi.org/10.5281/zenodo.22087646}. The corresponding GitHub repository is available at \url{https://github.com/epodlad/pds-source-gate-transfer}. The archive includes the frozen 12-event table, event-level phase and phase-jitter results, height-localization diagnostics, expansion tests, map/posterior arrays, and the scientific analysis scripts used for the reported diagnostics. Raw STEREO/SECCHI observations are publicly available from the mission archive and are not redistributed.

\begin{acknowledgments}

The author is sincerely grateful to the anonymous referee and the editor for their exceptionally careful and constructive work on the manuscript, especially for questions and suggestions that helped clarify which claims required stronger evidence and where the argument needed further strengthening.

The author thanks Angelos Vourlidas for discussions of periodic density structures and STEREO/SECCHI observations, Alexis P. Rouillard for discussions of possible PDS formation and release mechanisms, and Eva Robbrecht and Serge Koutchmy for discussions of coronagraphic measurements, streamer geometry, and their physical interpretation.

The author acknowledges the STEREO/SECCHI consortium and instrument teams. SECCHI was constructed by an international consortium led by the U.S. Naval Research Laboratory with contributions from institutions in the United States and Europe. This work uses public STEREO observations.

\end{acknowledgments}

\bibliographystyle{aasjournal}
\setlength{\bibsep}{0pt plus 0.2ex}
\bibliography{references}

\end{document}